\documentclass[titlepage,preprint,nofootinbib,prd,tightenlines,nopacs,
byrevtex,amsmath]{revtex4}

\usepackage{graphicx}
\usepackage{bm}
\newcommand{\beq}{\begin{equation}}
\newcommand{\eeq}{\end{equation}}
\newcommand{\eq}[1]{Eq.~(\ref{#1})}
\newcommand{\crossed}[1]{#1\hspace{-.5em}\slash}

\begin{document}

\preprint{UK/09-06}
\title {Radiative-Recoil Corrections to Hyperfine Splitting:\\ Polarization Insertions in the Muon Factor}
\author {Michael I. Eides}
\altaffiliation{Also at Petersburg Nuclear Physics Institute,
Gatchina, St.Petersburg 188300, Russia}
\email{eides@pa.uky.edu, eides@thd.pnpi.spb.ru}
\affiliation{Department of Physics and Astronomy,
University of Kentucky,
Lexington, KY 40506, USA,}

\author{Valery A. Shelyuto}
\email{shelyuto@vniim.ru}
\affiliation{D. I.  Mendeleev Institute of Metrology,
St.Petersburg 190005, Russia}

\begin{abstract}

We consider three-loop radiative-recoil corrections to hyperfine splitting in muonium due to insertions of one-loop polarization operator in the muon factor. The contribution produced by electron polarization insertions are enhanced by the large logarithm of the electron-muon mass ratio. We obtained all single-logarithmic and nonlogarithmic radiative-recoil corrections of order $\alpha^3(m/M)E_F$ generated by the diagrams with electron and muon polarization insertions.

\end{abstract}


\maketitle

\section{Introduction. Muon Factor Contribution to HFS\label{intro}}

\begin{figure}[tbh]
\includegraphics{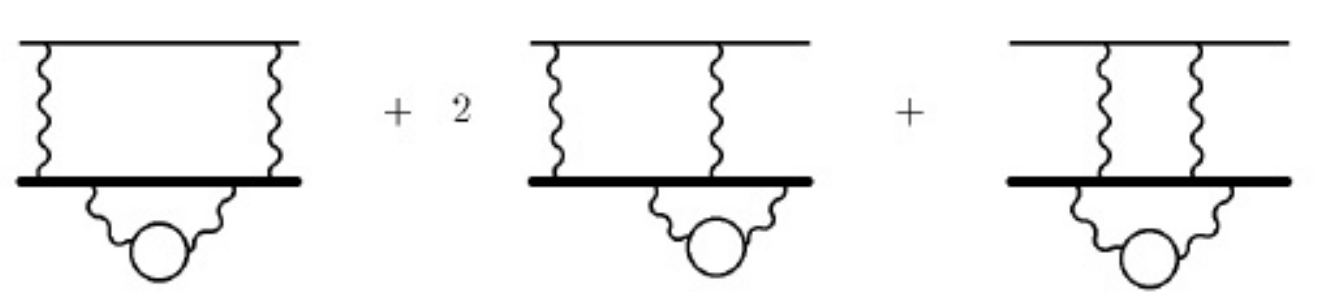}
\caption{\label{elpol} Electron polarization insertions}
\end{figure}

\begin{figure}[tbh]
\includegraphics{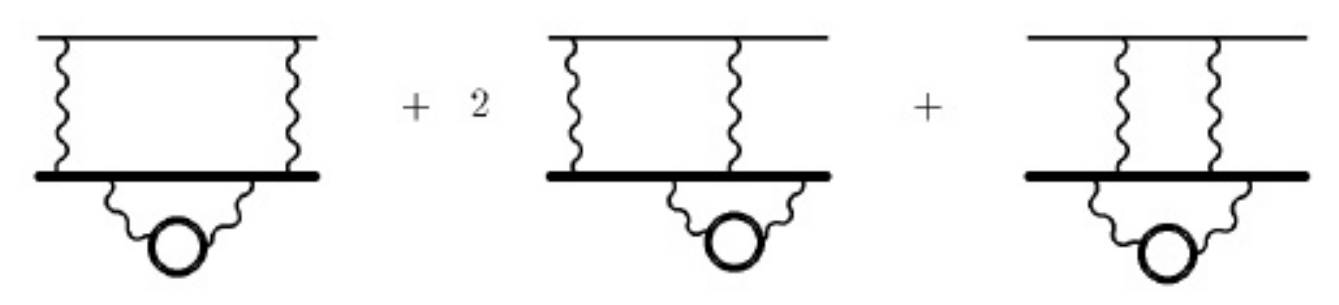}
\caption{\label{mupol} Muon polarization insertions}
\end{figure}

Calculation of numerous corrections to hyperfine splitting (HFS) in muonium has a long history (see, e.g., reviews in \cite{egsbook,preprts}). By now the largest challenge to the theory is calculation of three-loop radiative-recoil corrections. Below we consider such corrections generated by two sets of diagrams  with electron and muon vacuum polarization insertions in the radiative photon attached to the muon line in Fig.~\ref{elpol} and Fig.~\ref{mupol}.  The muon anomalous magnetic moment is subtracted from all vertices in these diagrams. Technically the gauge invariant anomalous magnetic moment is the hardest entry in the expression for the vertex at small transferred momenta, and this prompts its separation. It turns out that the anomalous magnetic moment does not generate radiative-recoil corrections (see, e.g., \cite{jetp94,eksann2}).

The scale of integration momenta in the graphs in Fig.~\ref{elpol} and Fig.~\ref{mupol} is determined by the muon mass, and is much higher than external virtualities determined by the characteristic atomic momenta. In order to obtain radiative-recoil corrections it is sufficient to calculate matrix elements of the diagrams in Fig.~\ref{elpol} and Fig.~\ref{mupol} with the mass-shell external momenta with vanishing spatial components. Then contribution to hyperfine splitting is given by the sum of matrix elements of the gauge invariant sets of diagrams in Fig.~\ref{elpol} and Fig.~\ref{mupol} calculated between the free electron and muon spinors and multiplied by the square of the Coulomb-Schr\"odinger wave function at the origin. Explicitly contributions to HFS generated by each of the gauge invariant  sets of graphs in Fig.~\ref{elpol} and Fig.~\ref{mupol} have the form

\beq \label{genexpress}
\Delta E=\frac{Z\alpha}{\pi}\frac{m}{M}E_F
\int \frac{d^4k}{\pi^2 i}\frac{1}{(k^2+ i0)^2}\langle\gamma^\mu\crossed k \gamma^\nu\rangle
\eeq
\[
\times\left(\frac{1}{k^2+4\mu k_0+i0}
+\frac{1}{k^2-4\mu k_0+i0}\right)L_{\mu\nu}
\equiv\Delta\epsilon\frac{Z\alpha}{\pi}\frac{m}{M}E_F,
\]

\noindent
where $m$ and $M$ are the electron and muon masses, respectively, $\mu=m/(2M)$, $k$ is dimensionless exchange photon momentum measured in terms of the muon mass, $\alpha$ is the fine structure constant, $E_F=(8/3)(Z\alpha)^4(m/M)m$ is the Fermi hyperfine splitting energy, $L_{\mu\nu}$ is the muon line factor, and the broken brackets denote matrix elements between hyperfine states. Constant $Z$ measures the muon charge in terms of the electron charge and is equal one in muonium, but we keep it as an indicator on the origin of different contributions.

\begin{figure}[tbh]
\includegraphics{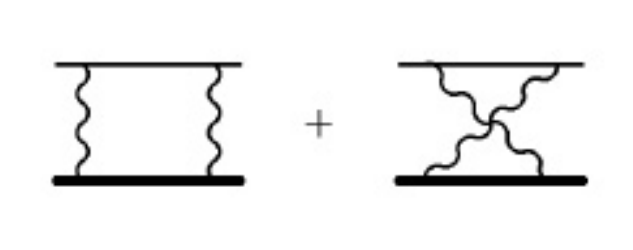}
\caption{\label{twoph} Two-photon exchange}
\end{figure}

\begin{figure}[tbh]
\includegraphics{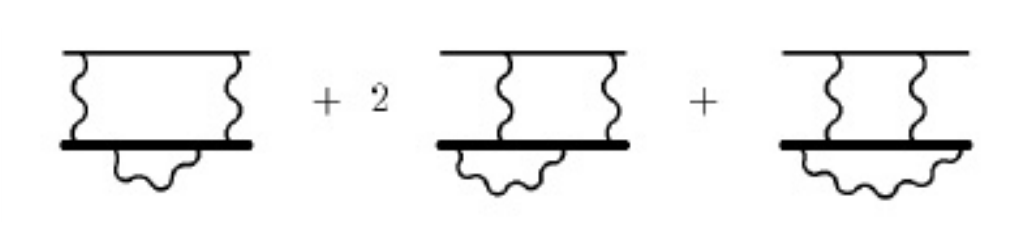}
\caption{\label{muoneloop} One-loop radiative insertions in the muon line}
\end{figure}

Generically dimensionless recoil contributions $\Delta\epsilon$ are enhanced by powers of large logarithm of the muon-electron mass ratio (see, e.g, \cite{egsbook,preprts}). Origin of these large logarithms is quite transparent. In the case of the skeleton muon factor $L_{\mu\nu}$  momentum integral in \eq{genexpress} is linearly divergent at low integration momenta of order $mZ\alpha$ (we temporarily return to dimensionful momenta in these considerations). This linear divergence corresponds to the classic nonrecoil Fermi contribution to HFS and should be subtracted in calculation of recoil contribution. After subtraction the skeleton integral in \eq{genexpress} becomes logarithmic in the wide integration region $m\leq k\leq M$ and generates the leading recoil correction to HFS. Considering one-loop radiative insertion in the muon line in Fig.~\ref{muoneloop} we need to separate gauge invariant one-loop anomalous magnetic moment and the remaining muon factor. The one-loop anomalous magnetic moment has the same momentum dependence as the skeleton contribution, and generates only nonrecoil contribution. Due to generalized low-energy theorem  for virtual Compton scattering \cite{yaf88,eksann2} the remaining one-loop muon factor is suppressed by an additional factor $k^2/M^2$ in comparison with the skeleton muon factor. This suppression not only makes the integral in \eq{genexpress} convergent in the infrared region, but also makes the integral nonlogarithmic. As a result one-loop muon factor generates only nonlogarithmic radiative-recoil corrections obtained in \cite{yaf88,jetp94,eks88,eksann2}. Insertion of vacuum polarization in the radiative photons does not change the low momenta behavior of the muon factor $L_{\mu\nu}$ and the integral over exchanged momenta in \eq{genexpress} corresponding to the diagrams in Fig.~\ref{elpol} and Fig.~\ref{mupol} remains convergent and nonlogarithmic. Then the leading recoil correction to HFS generated by the diagrams in Fig.~\ref{mupol} with the muon polarization insertions does not contain large logarithms and is a pure number. The case of of electron polarization insertions in Fig.~\ref{elpol} is a bit more complicated. While it is still true that integration over the exchanged momentum is nonlogarithmic, another source of large logarithms arises in these diagrams. As we already discussed all characteristic momenta in these diagrams are or order of the muon mass $M$. Then the electron polarization insertions in the diagrams in  Fig.~\ref{elpol} enter in the asymptotic regime, and the leading contribution to HFS generated by the diagrams with the electron polarization insertions in Fig.~\ref{elpol} is just the product of the leading asymptotic term in the high momentum expansion of the electron polarization operator and the radiative-recoil correction to HFS generated by the one-loop muon factor. In other words we can say that the leading logarithmic contribution to HFS generated by the diagrams in Fig.~\ref{elpol} is obtained from the respective nonlogarithmic contribution of the diagrams in Fig.~\ref{muoneloop} by substitution of the running coupling constant $\alpha(M)$ for radiative photons.

\section{Analytic Calculations}

Actual calculations start with consideration of separate contributions generated by the diagrams with self-energy, vertex, and spanning photon insertions in Fig.~\ref{elpol} and Fig.~\ref{mupol}. Redefining the muon factor by extracting from it all factors of $\alpha$, we write the general expression in \eq{genexpress} in the form

\beq
\Delta E=\Delta\epsilon\frac{\alpha(Z^2\alpha)(Z\alpha)}{\pi^3}\frac{m}{M}E_F,
\eeq

\noindent
where

\beq \label{threeterms}
\Delta\epsilon=\Delta\epsilon_\Sigma+2\Delta\epsilon_\Lambda+\Delta\epsilon_\Xi.
\eeq

\noindent
Explicit expressions for different terms on the right hand side in \eq{threeterms} are derived in the Feynman gauge. We start with the respective diagrams in Fig.~\ref{muoneloop} without polarization insertions but with massive radiative (but not exchanged) photons. We account for polarization insertions by considering photon mass squared to be $\lambda^2=16\mu^2/(1-v^2)$ or $\lambda^2=4/(1-v^2)$, for the electron and muon polarizations, respectively. Then insertion of the polarization operator means an additional integration over velocity $v$ with the weight

\beq \label{velintweight}
\int_0^1 \frac{{dv}}{1-v^2}~v^2\biggl(1-\frac{v^2}{3}\biggr).
\eeq

Below we collect integral representations for the contributions $\Delta\epsilon_i$ obtained directly from the Feynman diagrams. For the contribution of the self-energy diagram we obtained

\beq \label{selfrepr}
\Delta\epsilon_\Sigma=\frac{i}{2\pi^2}
\int_0^1 {dx} \int_0^x {dy}
\int_0^1 \frac{{dv}}{1-v^2}~v^2\biggl(1-\frac{v^2}{3}\biggr)
\int \frac{{d^4k}}{k^4}~\frac{2k^2}{k^4-16\mu^2 k_0^2}
\eeq
\[
\times\biggl\{3k_0 \frac{1-x^2}{\Delta_1}-(3k_0^2 - 2{\bm k}^2)
(1-x)^2\biggl[1-\frac{2(1+x)y}{\Delta_0}\biggr] \frac{1}{\Delta_1}~\biggr\}\equiv \Delta\epsilon_{\Sigma1}+\Delta\epsilon_{\Sigma2},
\]

\noindent
where

\[
\Delta_0(x)=x^2+\lambda^2(1-x)\equiv y(1-x)a_1^2(x,y),
\]
\[
\Delta_1(x,y)=y(1-x)(-k^2+2k_0+a_1^2-i0).
\]

The spanning photon contribution has the form

\beq  \label{spanrepr}
\Delta\epsilon_\Xi=-\frac{i}{4\pi^2}
\int_0^1 {dx} \int_0^x {dy}(x-y)
\int_0^1 \frac{{dv}}{1-v^2}~v^2\biggl(1-\frac{v^2}{3}\biggr)
\int \frac{{d^4k}}{k^4}\frac{2k^2}{k^4-16\mu^2 k_0^2}
\eeq
\[
\times\biggl\{2(3k_0^2 - 2{\bm k}^2)
\biggl[\frac{1-2y}{\Delta}+\frac{-2+2x(1-y)-x^2}{\Delta^2}
+\frac{-\lambda^2 (1-x)y}{\Delta^2}\biggr]
\]
\[
-6bk_0 \biggl[ \frac{1-2y}{\Delta}+\frac{2x(1-y)+x^2}{\Delta^2}
+\frac{\lambda^2 (1-x)(2-y)}{\Delta^2}+\frac{2bk_0 y(1-y)}{\Delta^2}
\biggr]\biggr\}\equiv \Delta\epsilon_{\Xi1}+\Delta\epsilon_{\Xi2},
\]

\noindent
where

\[
\Delta(x,y) = y(1-y)(-k^2 + 2bk_0 + a^2 -i0),
\]
\[
a^2(x, y) = \frac{x^2+\lambda^2 (1-x)}{y(1-y)},\quad
b(x, y) = \frac{1-x}{1-y}.
\]

The vertex contribution is as follows

\beq \label{vertrepr}
\Delta\epsilon_\Lambda=-\frac{i}{2\pi^2}
\int_0^1 {dx} \int_0^x {dy}
\int_0^1 \frac{{dv}}{1-v^2}v^2\biggl(1-\frac{v^2}{3}\biggr)
\int \frac{{d^4k}}{k^4}\frac{2k^2}{k^4-16\mu^2 k_0^2}
\eeq
\[
\times\biggl\{(3k_0^2 - 2{\bm k}^2)\biggl\{\biggl[-\bigl[(x-y)(1-2y)+ y(1-y)\bigr]\frac{1}{\Delta}
+2 \biggl(1-x-\frac{x^2}{2}\biggr)\frac{(1-x)y}{\Delta \Delta_0}
\]
\[
-x(1-x)\frac{(1-x)y}{\Delta \Delta_0}
+\frac{1}{\Delta}\bigl[1-x + (x-y)^2\bigr]\biggr]\biggr\}
-3k_0 \biggl[k_0 \frac{x(1-x)^2y}{\Delta  \Delta_0}
-\frac{1-x}{\Delta}
\biggr]
\]
\[
- \int_0^1 {du} \biggl\{ (3k_0^2 - 2{\bm k}^2)\biggl[
2 \biggl(1-x-\frac{x^2}{2}\biggr)\frac{k^2y^2(1-y)(x-y)}{\Delta_u^2 \Delta_0}
\]

\[
-x(1-x)\frac{k^2y^2(1-y)(x-y)}{\Delta_u^2 \Delta_0}
-\frac{k^2y(1-y)}{\Delta_u^2}\bigl[1-x + (x-y)^2\bigr]\biggr]
\]

\[
-3k_0^2   x(1-x)\frac{k^2y^2(1-y)(x-y)}{\Delta_u^2 \Delta_0}
\biggr\}\biggr\}\equiv \Delta\epsilon_{\Lambda1}+\Delta\epsilon_{\Lambda2}
+\Delta\epsilon_{\Lambda3},
\]

\noindent
where

\[
\Delta_u(x,y,u) = y(1-y) (-k^2 + 2b_uk_0 + a^2_u),
\]
\[
a_u^2(x,y,u) = ua^2(x,y),\qquad b_u(x,y,u) = \frac{1-y-xu+yu}{1-y} .
\]

Our next goal is to calculate the momenta integrals in \eq{selfrepr}, \eq{spanrepr}, and \eq{vertrepr} analytically. We explained in Section \ref{intro} that due to soft low-energy behavior of the muon factor $L_{\mu\nu}$ the momentum integral in \eq{genexpress} is convergent and nonlogarithmic. Hence, $\Delta\epsilon$ does not depend on the electron-muon ratio explicitly (it depends only on the effective photon mass $\lambda$), and one can safely omit factor $\mu$ in the electron denominators in the square brackets in \eq{genexpress}, what significantly simplifies all calculations. However, each of the exchanged momentum integrals in \eq{selfrepr}, \eq{spanrepr}, and \eq{vertrepr} corresponding to the self-energy, the spanning photon, and the vertex insertion contributions to the muon factor is logarithmic in the wide integration momenta region $m/M\leq k\leq 1$. At small integration momenta of order $m/M$ these would be divergences are cut off by the $\mu$-dependent electron denominators. This is the reason why we preserved parameter $\mu$ in sums of the electron propagators $2k^2/(k^4-16\mu^2k_0^2)$ in \eq{selfrepr}, \eq{spanrepr}, and \eq{vertrepr}. Separate contributions in \eq{selfrepr}, \eq{spanrepr}, and \eq{vertrepr} are gauge noninvariant, and one could hope to find a gauge where individual contributions are infrared finite even if one neglects the electron mass in the electron denominators. This is not the case, such gauge does not exist and infrared divergences survive in individual contributions even in the infrared soft Yennie gauge \cite{yaf88}.

We would like to get rid of the would be logarithmic divergencies at the level of integrands, so that divergent for $\mu=0$ momentum integrals would not arise at all. To this end we wrote all integrals in \eq{selfrepr}, \eq{spanrepr}, and \eq{vertrepr} in terms of similar Feynman parameters and organized them in three groups, combining contributions from different integrals. We obtained the representation for the energy splitting in the form

\beq
\Delta\epsilon=\Delta\epsilon_\Sigma+2\Delta\epsilon_\Lambda+\Delta\epsilon_\Xi
=\sum_{i=1}^{i=3}\Delta\epsilon_i,
\eeq

\noindent
where

\beq \label{1stgr}
\Delta\epsilon_1=\Delta\epsilon_{\Sigma2}+\Delta\epsilon_{\Xi1}
+2\Delta\epsilon_{\Lambda1},
\eeq
\beq \label{2ndgr}
\Delta\epsilon_2=\Delta\epsilon_{\Sigma1}+\Delta\epsilon_{\Xi2}
+2\Delta\epsilon_{\Lambda2},
\eeq
\beq \label{3dgr}
\Delta\epsilon_3=2\Delta\epsilon_{\Lambda3}.
\eeq

\noindent
Momentum integrals in \eq{1stgr} and \eq{2ndgr} are apparently logarithmically divergent at $\mu=0$, and require additional transformations.

Consider first $\Delta\epsilon_1$ in \eq{1stgr}. After some integrations by parts over the Feynman parameters it acquires the form

\beq
\Delta\epsilon_1=\frac{i}{\pi^2}\int_0^1 {dx} \int_0^x {dy}
\int_0^1 \frac{{dv}}{1-v^2}v^2\biggl(1-\frac{v^2}{3}\biggr)
\int \frac{{d^4k}}{k^2(k^4-16\mu^2 k_0^2)}(3k_0^2 - 2{\bm k}^2)\biggl(\frac{A_1}{\Delta}+ \frac{k^2B_1}{\Delta^2}
\biggr),
\eeq

\noindent
where

\[
A_1 (x, y, \lambda)=
3\biggl(1-\frac{2xy}{\Delta_0}\biggr)-5x+3x^2 +(8-4x)y +\biggl(2-\frac{4x}{\Delta_0}\biggr)y^2
+ \frac{\lambda^2 (1-x)y}{\Delta_0}(-9+3x-2y) ,
\]

\noindent
and

\[
B_1 = \biggl[x \biggl(1-\frac{x}{2}\biggr)-2\biggl(1+\frac{x}{2}\biggr)
\frac{x(1-x)y}{\Delta_0}\biggr]\biggl[ y(1-y) + (-2+x)\frac{xy^2}{\Delta_0} \biggr]
\]
\[
+\bigl[-2+2x(1-y)-x^2-\lambda^2(1-x)y\bigr] \frac{y^2(x-y)}{\Delta_0}.
\]

\noindent
After the Wick rotation the integral for $\Delta\epsilon_1$ in four-dimensional Euclidean spherical coordinates ($k_0=k\cos\theta$, $|{\bm k}|=k\sin\theta$) has the form

\beq \label{dele1mu}
\Delta\epsilon_1=\int_0^1 {dx} \int_0^x {dy}\int_0^1
\frac{{dv}}{1-v^2}v^2\biggl(1-\frac{v^2}{3}\biggr)
\int_0^{\infty} \frac{k^2dk^2}{k^4+16\mu^2k^2\cos\theta^2}\frac{2}{\pi} \int_0^{\pi} {d \theta}  \sin^2{\theta}(2 + \cos^2{\theta})
\eeq
\[
\times\biggl[\frac{A_1}{y(1-y)}+ \frac{k^2B_1}{y^2(1-y)^2}\frac{\partial}{\partial a^2}\biggr]
\frac{k^2+a^2}{(k^2+a^2)^2+4b^2k^2\cos^2{\theta}}.
\]

\noindent
Now we see that omitting the term with $\mu^2$ in the denominator would result in an apparently logarithmically divergent integral. However, this is not the case since the coefficient before the would be divergence turns into zero due to the identity

\beq \label{indnt1}
\int_0^1 {dx} \int_0^x {dy}\frac{A_1(x, y, \lambda)}{\Delta_0}=0,
\eeq

\noindent
that is valid for any $\lambda$. This means that the momentum integral in \eq{dele1mu} is convergent even at $\mu=0$. We obtain an explicitly convergent form for this integral by subtraction from the integrand taken at $\mu=0$ its leading low momentum asymptotic term. This subtraction does not change the integral since the coefficient before the the subtracted term is identically zero due to the identity in \eq{indnt1}. Of course, we have chosen this particular combination of  terms on the right hand side in \eq{1stgr} precisely  because we wanted to organize an infrared finite contribution. After subtraction and momentum integration we obtain a finite integral representation for $\Delta\epsilon_1$

\beq \label{eps1final}
\Delta\epsilon_1=\int_0^1 {dx} \int_0^x {dy}\int_0^1
\frac{{dv}}{1-v^2}v^2\biggl(1-\frac{v^2}{3}\biggr)
 \biggl\{-\frac94 \frac{A_1}{\Delta_0} \ln{\frac{1}{a^2}}
\eeq
\[
-\frac34 A_1 \biggl[\frac{3}{\Delta_0}L_0
+ \frac{1-y}{y(1-x)^2}\bigl(2L_0 + L_1\bigr) \biggr]
-\frac32  \frac{B_1}{y^2(1-x)^2}\bigl(L_0 + L_1 \bigr)\biggr\},
\]

\noindent
where $L_n =b^2\int_0^1{ds}s^n/(a^2+b^2s)$, or explicitly

\beq
L_0=\ln{\frac{a^2+b^2}{a^2}},\qquad L_1 =1- \frac{a^2}{b^2}L_0.
\eeq

Consider now the contribution in \eq{2ndgr} that is defined by the integral

\beq
\Delta\epsilon_2=\frac{3i}{2\pi^2}
\int_0^1 {dx} \int_0^x {dy}
\int_0^1 \frac{{dv}}{1-v^2}v^2\biggl(1-\frac{v^2}{3}\biggr)
\int \frac{{d^4k}}{k^4}\frac{2k^2}{k^4-16\mu^2 k_0^2}
\eeq
\[
\times k_0 \biggl[ \frac{x(1-x)}{\Delta_0}+\frac{ A_2}{\Delta}
+\frac{k^2x(1-x)y(1-y)}{\Delta \Delta_0}+ \frac{k^2B_{2}}{\Delta^2} \biggr],
\]

\noindent
where

\[
A_2(x,y,\lambda)=(1-x)(2x-y)+(-2+x)\frac{x^2y}{\Delta_0},
\]
\[
B_{2} =y \biggl(3x - 2y - \frac{x^2}{2} - \frac{3x^2y}{2} + xy^2\biggr)
+(-2+x)\biggl(1+\frac{3x}{2}-y\biggr)\frac{x^2y^2}{\Delta_0}.
\]

\noindent
After the Wick rotation we obtain in four-dimensional Euclidean spherical coordinates an integral

\beq  \label{dele2mu}
\Delta\epsilon_2=-3\int_0^1 {dx} \int_0^x {dy}
\int_0^1 \frac{{dv}}{1-v^2}v^2\biggl(1-\frac{v^2}{3}\biggr)2b\int_0^{\infty} \frac{k^4dk^2}{k^4+16\mu^2k^2\cos\theta^2}
\eeq
\[
\times\frac{2}{\pi} \int_0^\pi {d \theta}\sin^2{\theta}\cos^2{\theta}
\biggl[-\frac{A_2}{y(1-y)k^2}
+\frac{x(1-x)}{\Delta_0}
-\frac{B_{2}}{y^2(1-y)^2}\frac{\partial}{\partial a^2}\biggr]
\frac{1}{(k^2+a^2)^2+4b^2k^2\cos^2{\theta}},
\]

\noindent
that is apparently logarithmically divergent for $\mu=0$. Like in the case of the contribution $\Delta\epsilon_1$ the coefficient before the would be divergence turns into zero due to the identity

\beq \label{indnt2}
\int_0^1 {dx} \int_0^x {dy}\frac{(1-x)y}{\Delta_0^2} A_2(x,y,\lambda)=0,
\eeq

\noindent
that is valid for any $\lambda$. Using this identity we subtract the would be divergent at $\mu=0$ integral of the leading low momentum asymptotic term in the integrand in \eq{dele2mu}, and obtain a finite integral representation for $\Delta\epsilon_2$

\beq  \label{eps2final}
\Delta\epsilon_2=-3\int_0^1 {dx} \int_0^x {dy}
\int_0^1 \frac{{dv}}{1-v^2}v^2\biggl(1-\frac{v^2}{3}\biggr)
\biggl[\frac{(1-x)yA_2}{2\Delta_0^2} \biggl(\ln{\frac{1}{a^2}}+L_0\biggr)
\eeq
\[
+\frac{A_2}{2b\Delta_0} L_1
+\frac{x(1-y)}{\Delta_0} (L_0-L_1)
+\frac{B_{2}}{y(1-x)\Delta_0} L_1\biggr].
\]

The momentum integral in \eq{3dgr} is convergent even at $\mu=0$, and since we are interested in the leading recoil contribution we calculate it in this limit
\beq   \label{eps3final}
\Delta\epsilon_3  =
3\int_0^1 {dx} \int_0^x \frac{{dy}}{y(1-y)}
\int_0^1 \frac{{dv}}{1-v^2}v^2\biggl(1-\frac{v^2}{3}\biggr)\int_0^1 \frac{{du}}{b_u^2}
\eeq
\[
\times\biggl\{(L_{0u}+L_{1u}) \biggl[(2-3x)\frac{y(x-y)}{\Delta_0}
-\bigl[1-x + (x-y)^2\bigr]\biggr]
+(L_{0u}-3L_{1u}) x(1-x)\frac{y(x-y)}{\Delta_0}\biggr\}.
\]

\noindent
where

\beq
L_{0u}=\ln{\frac{a_u^2+b_u^2}{a_u^2}},\qquad L_{1u} =1- \frac{a_u^2}{b_u^2}L_{0u}.
\eeq

\section{Numerical Calculations and Results}

Let us notice that the expressions for $\Delta\epsilon_i$ obtained in \eq{eps1final}, \eq{eps2final}, and \eq{eps3final} allow us to reproduce the radiative-recoil correction to HFS generated by the diagrams in Fig.~\ref{muoneloop} with one-loop radiative insertions in the muon line. To obtain these corrections we need to omit integration over velocity $v$ and all explicit factors containing $v$ in \eq{eps1final}, \eq{eps2final}, and \eq{eps3final}, and also let the radiative photon mass $\lambda$ to be zero. The resulting integrals admit analytic calculation and the results are collected in Table~\ref{oldres}. Restoring the overall dimensional factor we obtain

\beq \label{oldrees}
\Delta E=\Delta\epsilon^{(0)}\frac{(Z^2\alpha)(Z\alpha)}{\pi^2}\frac{m}{M}E_F
=\biggl(\frac{9}{2} \zeta (3) - 3\pi^2\ln{2}+\frac{39}{8}\biggr)
\frac{(Z^2\alpha)(Z\alpha)}{\pi^2}\frac{m}{M}E_F,
\eeq

\noindent
what precisely reproduces the old result \cite{eks88,jetp94}.

\begin{table}
\caption{One-loop radiative recoil corrections\label{oldres}}
\begin{ruledtabular}
\begin{tabular}{l|c|}
$\Delta\epsilon_1^{(0)}$&$-15 \zeta{(3)} + 6\pi^2 \ln{2} - 5\pi^2 + \frac{237}{8}$ \\

$\Delta\epsilon_2^{(0)}$ &$\frac{15}{8}\zeta (3) - \frac{3\pi^2}{4}\ln{2} + \frac{\pi^2}{4}- \frac34$ \\

$\Delta\epsilon_3^{(0)}$ &$\frac{141}{8} \zeta (3) - \frac{33\pi^2}{4}\ln{2}+ \frac{19\pi^2}{4}-24$\\

\hline
$\Delta\epsilon^{(0)}$&$\frac{9}{2} \zeta (3) - 3\pi^2\ln{2}+\frac{39}{8}$  \\

\end{tabular}
\end{ruledtabular}
\end{table}

Now we are ready to proceed to calculations of the contributions to HFS generated by the diagrams in Fig.~\ref{elpol} and Fig.~\ref{mupol}. The leading logarithmic term corresponding to the diagrams in Fig.~\ref{elpol} with the electron polarization insertions in the radiative photon is obtained by substituting the running coupling constant $\alpha(M)$ for the radiative photon in the result in \eq{oldrees}, and is equal to (we do not include $\alpha$ in the definition of $\Delta\epsilon$, it will be restored in the finite results)

\beq
\Delta\epsilon^{\log}=\frac{2}{3}\ln\frac{M}{m}\Delta\epsilon^{(0)}.
\eeq

\noindent
Similar equations are also valid for each of the contributions $\Delta\epsilon_i^{(0)}$. For practical calculations we defined nonlogarithmic contributions to HFS as

\beq \label{cidef}
\Delta C_i=\Delta\epsilon_i-\biggl(\frac{2}{3}\ln\frac{M}{m}-\frac{5}{9}\biggr)
\Delta\epsilon_i^{(0)}.
\eeq

\noindent
We calculated contributions in \eq{cidef} numerically using the integral representations in \eq{eps1final}, \eq{eps2final}, and \eq{eps3final}, with the photon mass squared $\lambda^2=16\mu^2/(1-v^2)$, and obtained

\beq
\Delta C_1=-1.4965(7),\qquad \Delta C_2=2.2442(6),\qquad \Delta C_{3}=5.7905(5).
\eeq

\noindent
Then the total contribution to HFS due to electron polarization insertions in Fig.~\ref{elpol} has the form

\beq \label{epse}
\Delta \epsilon_{e}=\biggl(3\zeta (3) - 2\pi^2\ln{2}+\frac{13}{4}\biggr)
\ln{\frac{M}{m}} + 12.227(2).
\eeq

The contribution to HFS due to muon polarization insertions in Fig.~\ref{mupol} was calculated numerically with the help of the integral representations in \eq{eps1final}, \eq{eps2final}, and \eq{eps3final}.  This time $\lambda^2=4/(1-v^2)$ plays the role of the photon mass squared in the integrands, and we obtained

\beq
\Delta\epsilon_1 =0.2987,\qquad \Delta\epsilon_2 =-0.0003,\qquad \Delta\epsilon_3 =-1.2291,
\eeq

\noindent
or

\beq  \label{epsmu}
\Delta\epsilon_\mu=-0.9307.
\eeq

\noindent
Restoring all dimensional factors we obtain from \eq{epse} and \eq{epsmu}

\beq
\Delta E_{e}=\biggl[
\biggl(3\zeta (3) - 2\pi^2\ln{2}+\frac{13}{4}\biggr)
 \ln{\frac{M}{m}} + 12.227(2)\biggr]
\frac{\alpha(Z^2\alpha)(Z\alpha)}{\pi^3}\frac{m}{M}E_F,
\eeq
\beq
\Delta E_\mu
=-0.931\frac{(Z^2\alpha)^2(Z\alpha)}{\pi^3}\frac{m}{M}E_F.
\eeq

\noindent
Total contribution to HFS due to diagrams with electron and muon polarization insertions  in Fig.~\ref{elpol} and Fig.~\ref{mupol} is

\beq
\Delta E=\biggl[
\biggl(3\zeta (3) - 2\pi^2\ln{2}+\frac{13}{4}\biggr)
 \ln{\frac{M}{m}} + 11.297(2)\biggr]
\frac{\alpha^3}{\pi^3}\frac{m}{M}E_F.
\eeq

This result together with the results \cite{egs01,egs03,egs04,es2009} completes calculation of all radiative-recoil corrections of order $\alpha^3(m/M)E_F$ generated by the diagrams with  electron and muon polarization insertions.

\begin{acknowledgments}
This work was supported by the NSF grant PHY-0757928. V.A.S. was also
supported in part by the RFBR grants 06-02-16156 and 08-02-13516, and by the DFG grant GZ  436 RUS 113/769/0-3.
\end{acknowledgments}

\end{document}